\documentclass[twocolumn,aps,prb,superscriptaddress,floatfix]{revtex4-1}
\usepackage{graphicx}
\usepackage{dcolumn}
\usepackage[T1]{fontenc}
\usepackage{amsmath}
\usepackage{sidecap}

\begin{document}
\title{Microwave magnetic field modulation of spin torque oscillator based on perpendicular magnetic tunnel junctions}

\author{Witold Skowro\'{n}ski}
 \email{skowron@agh.edu.pl}
\author{Jakub Ch\k{e}ci\'{n}ski}
\author{S\l{}awomir Zi\k{e}tek}
\affiliation{AGH University of Science and Technology, Department of Electronics, Al. Mickiewicza 30, 30-059 Krak\'{o}w, Poland}
%\author{Hitoshi Kubota}
\author{Kay Yakushiji}
\author{Shinji Yuasa}
\affiliation{National Institute of Advanced Industrial Science
and Technology, Spintronics Research Center, Tsukuba, Ibaraki 305-8568, Japan}

\begin{abstract}
Modulation of a spin-torque oscillator (STO) signal based on a magnetic tunnel junction (MTJ) with perpendicularly magnetized free layer is investigated. Magnetic field inductive loop was created during MTJ fabrication process, which enables microwave field application during STO operation. The frequency modulation by the microwave magnetic field of up to 3 GHz is explored, showing a potential for application in high-data-rate communication technologies. Moreover, an inductive loop is used for self-synchronization of the STO signal, which after field-locking exhibits significant improvement of the linewidth and oscillation power.
\end{abstract}

\maketitle

\section{Introduction}
\label{sec:intro}
Spin transfer torque nano oscillators based on both fully-metallic multilayers \cite{tsoi_generation_2000} and magnetic tunnel junctions (MTJ) \cite{deac_bias-driven_2008} are widely studied due to their potential applications in future wireless communication \cite{chen_sto_review_2015} and magnetic recording \cite{Braganca_nanoscale_2010}. Up to now, the most promising parameters in terms of oscillation power and linewidth have been measured for MTJs with an in-plane magnetized reference layer and perpendicularly magnetized free layer \cite{Taniguchi_critical_2013, maehara_large_2013}, which also enable magnetic-field-free operation \cite{skowronski_zero-field_2012, Zeng2013}. %It was also shown that such magnetic configuration is optimal for synchronization of two MTJ connected \cite{Taniguchi_mutual_2017}. 
To further enhance spin torque oscillator (STO) characteristics,  phase-locking to another STO \cite{kaka_mutual_2005} or external microwave signals \cite{singh_self_2018} were suggested. Recently, electronic phase-lock loop (PLL) using uniform precession mode oscillations \cite{tamaru_extremely_2015} and magnetic vortex \cite{kreissig_vortex_2017} have been presented. In addition, self-synchronization via a field loop was shown in the in-plane magnetized MTJs \cite{Kumar2016}.

Another important feature of STO is the ability to tune the oscillation frequency over wide range, which is essential for wireless communication devices. Precession frequency of operating STO depends on the bias current due to its inherent nonlinear nature \cite{slavin_nonlinear_2009}, which is used in the aforementioned PLL circuits. Various modulation techniques presented in uniformly precessing MTJs \cite{ruiz-calaforra_frequency_2017, muduli_nonlinear_2010} and vortex-based oscillators \cite{manfrini_frequency_2011} were shown; however, an even stronger dependence of the STO frequency on an external magnetic field is expected, which would allow for higher modulations rates, as numerically predicted in Ref. \cite{purbawati_enhanced_2016}. MTJs with an additional magnetic field line were already tested for STO-based read head sensors with a transition time down to 2-ns \cite{nagasawa_frequency_2011}.

In this work, we present an experimental study of STO based on an MTJ with a perpendicularly magnetized free layer. In contrast to the vortex based STO \cite{tsunegi_self_2016}, we use a uniform oscillation mode of the free layer, which is characterized by higher oscillation frequencies. An additional, isolated microwave line was fabricated on the top of the MTJ, which enables microwave magnetic field application during STO operation. At optimal configuration (static magnetic field, current bias), the STO operation frequency of $f$ = 5 GHz and linewidth $\delta f$ = 10 MHz were measured, which enables field-modulation at the rate of up to 3 GHz. Moreover, after injecting the amplified STO signal to the field line, and creating so-called inductive feedback loop, we observed the quality factor improvement to $Q$ > 18000; however, at the cost of bringing the system to the edge of stability.

\section{Experiment}
\label{sec:experiment}
\begin{figure}[t!]
\centering
\includegraphics[width=\columnwidth]{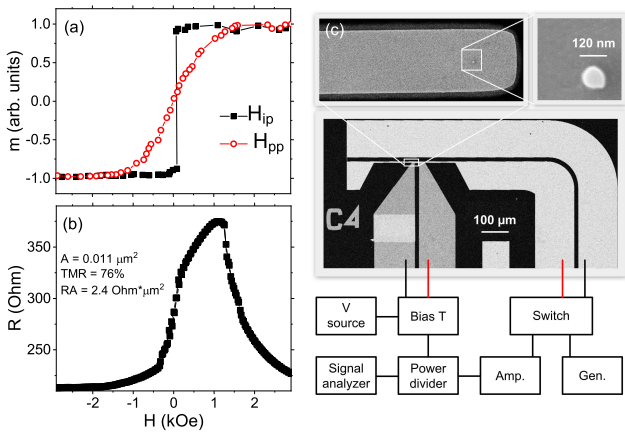}
\caption{(a) magnetization loops measured in the magnetic field applied in the sample plane $H_{ip}$ and perpendicular to the sample plane $H_{pp}$, (b) resistance vs. in-plane magnetic field loop of a fabricated MTJ, (c) micrograph of the device (the top two images were recorded during the fabrication process) with a schematic representation of the measurement setup.}
\label{fig:fig1}
\end{figure}

Multilayer of the following structure: Ru(4)/ IrMn(6)/ CoFe(2.5)/ Ru(0.8)/ CoFeB(2.5)/ MgO (1)/ CoFeB(1.8)/ MgO(1)/ Ru(3)/ Ta(5)/ Ru(2)/ Pt(3) (thickness in nm) was deposited on chemically-mechanically polished Si wafer using magnetron sputtering method. The bottom CoFeB is pinned to the synthetic antiferromagnetic structure and acts as an in-plane magnetized reference layer. The magnetic properties of the films were determined using vibrating sample magnetometry (VSM). MTJ device was fabricated using electron beam lithography, ion-beam milling and lift-off processes into circular and elliptical nanopillars of diameters ranging from 120 to 450 nm, with the additional field line, separated from the top and bottom contact by a 100-nm thick Al$_2$O$_3$ layer. Electrical contacts and the field line was made of Al(20)/Au(30) conducting layer. The micrograph of the MTJ design is presented in Fig. \ref{fig:fig1}(c). % After the nanofabrication process, MTJ revealed an out-of-plane easy axis of the free layer, which is manifested by the hysteresis-free and gradual resistance vs. in-plane magnetic field dependence, presented in Fig. \ref{fig:fig1}(b).

Electrical transport properties were determined in the rotating probe station enabling magnetic field application at an arbitrary angle with respect to the MTJ axis. Custom design microwave, five-tip probe (T-Plus) was used to connect a bias source (Keithley 2401) and signal analyser (Agilent PXA N9030A) via broadband bias-T (Mini Circuits ZX86-12G) to the MTJ. An external radio-frequency (rf) signal (Agilent E8257D) was fed to the field line using the same microwave probe. Alternatively, the amplified (Mini circuits ZVA-213, 26 dB Gain) STO signal was connected to field line via broadband power divider (Mini Circuits ZN2PD2-63). Transport and rf properties were determined in ten different devices exhibiting qualitatively similar results.  

In addition, static transport measurements were performed in a perpendicular magnetic field using the four-point method on different MTJs fabricated on the same wafer, in order to extract the resistance area (RA) product and the tunnelling magnetoresistance (TMR) ratio.

\section{Results and discussion}
\label{sec:results}
VSM measurements are presented in Fig. \ref{fig:fig1}(a). Due to a significant contribution of a double CoFeB/MgO interface to the magnetic anisotropy of the top free layer, the saturation field value in perpendicular orientation is $H_\mathrm{K}$ = 1.2 kOe, indicating effectively weak in-plane magnetic anisotropy. After the nanofabrication process, the demagnetizing field in the perpendicular direction decreases, which leads to an effective perpendicular anisotropy of the free layer of the patterned device - Fig. \ref{fig:fig1}(b). The TMR ratio reaches 76\% with the RA product of 2 Ohm$\times \mu$m$^2$. 

\begin{figure}[t!]
\centering
\includegraphics[width=\columnwidth]{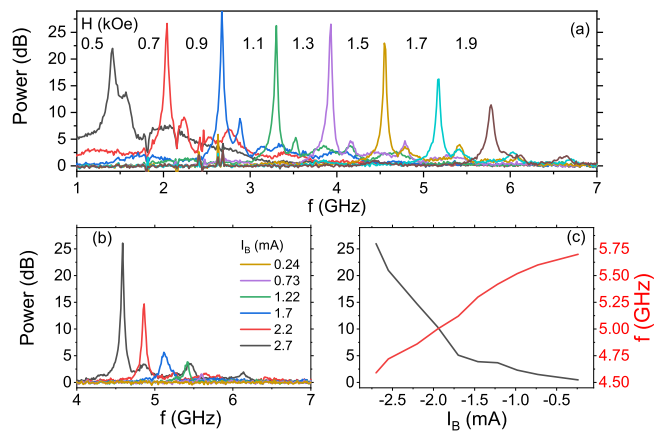}
\caption{(a) STO power (expressed in dB over the noise floor) measured in different magnetic fields applied at 65$^\circ$ with respect to the sample plane, (b) bias current dependence on the STO signal. The STO frequency and power extracted from (b) are presented as a function of the bias current (c). }
\label{fig:fig2}
\end{figure}

After application of a negative bias voltage, indicating electron flow from the top free layer to the bottom reference layer, i.e., destabilization of the parallel magnetization alignment, the STO signal is measured in the presence of an external magnetic field. An example of the auto-oscillation signal measured in a magnetic field applied at $\theta$ = 65$^\circ$ with respect to the sample plane is presented in Fig. \ref{fig:fig2}(a). For selected amplitudes of magnetic fields, the linewidth at half maximum drops to around 10 MHz, resulting in a quality factor of several hundreds. We note that the STO signal is present for a range of $\theta$ between 50 and 90$^\circ$, which varies slightly between devices. An example STO signal vs. bias current dependence is presented in Fig. \ref{fig:fig2}(b), together with the STO frequency and power (expressed in dB over noise floor) - Fig. \ref{fig:fig2}(c). Reversing the bias polarity results in a much weaker STO signal and broader linewidth (not shown), similarly to other reports \cite{Kowalska_tunnel_2019}. 

\begin{figure}[t!]
\centering
\includegraphics[width=\columnwidth]{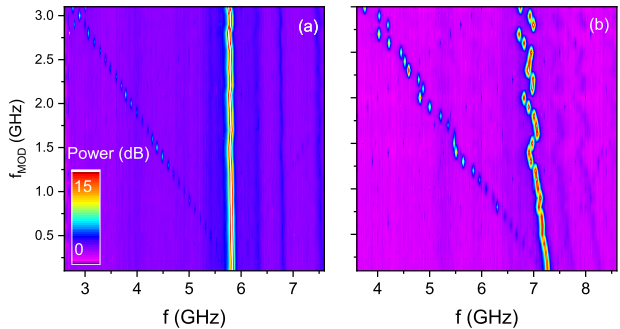}
\caption{STO signal with field-modulation of $f_\mathrm{MOD}$ up to 3 GHz and $P_\mathrm{MOD}$ = 0 dBm (a) and $P_\mathrm{MOD}$ = 10 dBm (b). Deviation from the main STO peak is present for higher modulation power.}
\label{fig:fig3}
\end{figure}

Next, we explore the magnetic field modulation possibilities of the STO signal. The magnetic field was kept constant $H$ = 2 kOe at $\theta$ = 65$^\circ$, resulting in auto-oscillations at around 5.7 GHz. After applying an rf signal of frequency $f_\mathrm{MOD}$ to the magnetic field line, modulation sidebands are clearly visible; however, only the frequencies below the main peak - Fig. \ref{fig:fig3}(a). The modulation power was set to$P_\mathrm{MOD}$ = 0 dBm. Single sideband modulation has been recently observed in a current-driven STO and such behaviour was explained by the non-linear dependence of the frequency and power on bias voltage \cite{Sharma_high-speed_2017}. Strikingly, in our case a clear modulation signal was observed up to $f_\mathrm{MOD}$ = 3 GHz, which exceeds frequency modulation range reported to date in the in-plane magnetized MTJs.

%To further elucidate on the field-modulated STO 
%In order to exclude a direct bias modulation (i.e. not via the Oersted field) that may induce in the supply bias line, 
To further elucidate on the microwave field-modulated STO behaviour, we repeated the frequency spectrum measurements for increased $P_\mathrm{MOD}$. Figs. \ref{fig:fig3}(b) present the $f_\mathrm{MOD}$ vs. frequency map for increased $P_\mathrm{MOD}$ = 10 dBm. In contrast to the measurement performed at $P_\mathrm{MOD}$ = 0, where a linear dependence of the sideband and modulation frequency is measured, an increased modulation power results in the main peak frequency deviation from its original position, determined by the bias current and magnetic field. We ascribe this effect to the injection-locking of the STO to higher harmonics of the modulating signal \cite{urazdin_fractional_2010, singh_integer_2017}. 

\begin{figure}[t!]
\centering
\includegraphics[width=\columnwidth]{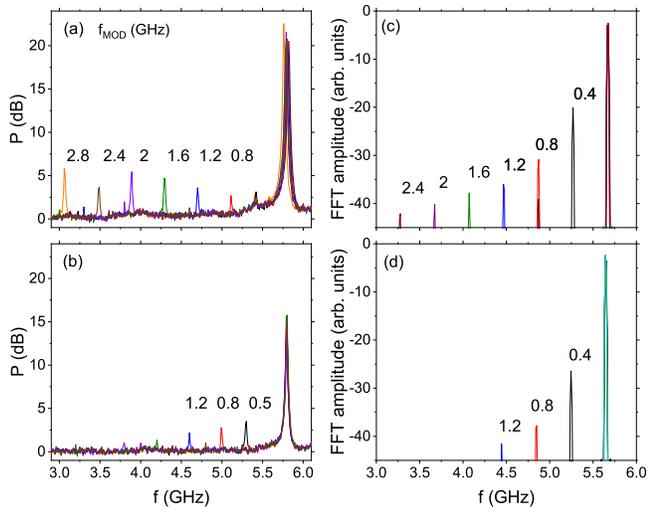}
\caption{Comparison between magnetic field modulation (a) and current bias modulation (b) indicating higher modulation rate for Oersted field method, as predicted in \cite{ruiz-calaforra_frequency_2017}. Corresponding macrospin simulation results are presented in (c) and (d).}
\label{fig:fig4}
\end{figure}

In order to exclude possible modulation mechanism via bias current that is induced in a neighbouring supply line from the magnetic field line, the modulation signal was also connected directly to the bias-tee via radio-frequency combiner in a separate experiment. In this case, the modulation signal was also measured, but only up to $f_\mathrm{MOD}$ = 1.2 GHz, which is well below the field-modulated range that exceeds 3 GHz -  Figs. \ref{fig:fig4}(a-b).

%In order to reproduce the observed effect theoretically, the macrospin simulation were performed.... (results may be included to Fig. 4.)

To reproduce the observed effect theoretically, macrospin simulation were performed. The parameters of the system were chosen to follow the experimental values whenever possible and were as following: $\mu_0 M_S$ = 1.6 T, maximum effective anisotropy field 300 Oe, damping constant $\alpha$ = 0.01, junction diameter 120 nm, free layer thickness 1.8 nm. The reference layer magnetization was assumed to be fixed in the sample plane and the small effective dipolar field originating from bottom parts of the stack was included in the free layer. The Landau-Lifshitz-Gilbert-Slonczewski equation was integrated numerically, assuming an angular dependence of the spin torque term with the angular term $\lambda$ \cite{taniguchi2013critical, kubota2013spin} equal to approximately 0.69. By applying a bias current of -2.05 mA and an external magnetic field of 2.2 kOe at $\theta$ = 85$^\circ$, we were able to obtain stable oscillations at a frequency similar to the one presented in figure \ref{fig:fig4}(a-b). Next, we performed a set of simulations for two different modulation approaches: one driven by the external magnetic field - \ref{fig:fig4}(c), and one driven by the bias current - \ref{fig:fig4}(d). Clearly, the modulation rate is significantly stronger in the case of the magnetic field modulation, further supporting the conclusion obtained from the experiment.

\begin{figure}[t!]
\centering
\includegraphics[width=\columnwidth]{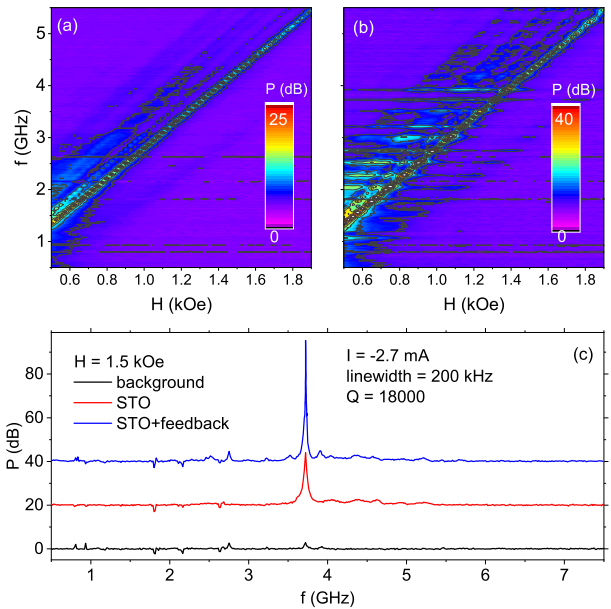}
\caption{STO performance with an inductive feedback-loop. Oscillation frequency vs. magnetic field applied in the $\theta$ = 65$^\circ$ recorded without (a) and with (b) the feedback loop connected. Horizontal lines in (b) represents the inverse of feedback loop delay of around 0.2 GHz. (c) example STO plots with the feedback loop turned off and on together with a background signal. Curves are offset by 20 dB for clarity.}
\label{fig:fig5}
\end{figure}

Finally, auto-synchronization of the STO with a feedback loop was explored. The microwave STO signal was amplified and fed to the inductive magnetic field line. The dependences of the STO power on the magnetic field without and with the feedback loop are presented in Fig. \ref{fig:fig5}(a-b). Without a feedback connected, the STO power reaches 25 dB over noise and monotonically decreases with increasing magnetic field. In the case of the feedback-STO, apart from the main peak, several side-peaks are visible in each spectrum. The distance (in the frequency domain) between peaks is fixed to around 0.2 GHz, which is a result of delay caused by the total feedback line length of around 1.5 m. The measured amplitude of each peak increases when the free-running oscillation frequency at a given bias and magnetic field corresponds to the resonance frequency of the feedback loop. This results in an increased oscillations power of up to 30 dB with respect to the STO without a feedback. An example showing one of the strongest STO signals measured with the feedback loop is presented in \ref{fig:fig5}(c). %We note that a small peak is present also when the bias voltage is well below the oscillation threshold. 
We note that the positive feedback used in the circuit design may  cause the system to be at the stability limit, i.e., further increase in gain of the amplifier may case self-excitation of the circuit even without STO presence. Nevertheless, the obtained STO Q factor of above 18000 and the oscillation power spectral density reaching 25 nV/$\sqrt{Hz}$ are promising for future applications.

\section{Summary}
\label{sec:summary}
In conclusion, the STO properties of the magnetic tunnel junction with perpendicularly magnetized free layer were investigated in the presence of an additional microwave magnetic field. Application of the external field enabled STO frequency modulation up to 3 GHz, which exceeds the limit for the bias-current modulation methods. By connecting the inductive field line to the amplified STO signal and creating a feedback loop, greater power and a Q-factor of up to 18000 were obtained.

\section*{Acknowledgments}
The research leading to these results has received funding from the Polish National Centre for Research and Development under grant No. LIDER/467/L-6/14/NCBR/2015. Nanofabrication was performed at Academic Centre for Materials and Nanotechnology of AGH University. Numerical calculations were supported in part by PL-GRID infrastructure.

%\begin{thebibliography}{10}
%\end{thebibliography}
\bibliographystyle{apl}
\bibliography{Skowronski_library}
\end{document}